\documentclass[journal]{IEEEtran}

\newif\ifIEEE
\IEEEtrue

\usepackage[nolist]{acronym}
\usepackage{cite}
\usepackage{color}
\usepackage{mathrsfs}
\usepackage{graphicx}
\usepackage{subfig}
\usepackage{amsmath}
\usepackage{amsfonts}
\usepackage{amssymb}
\usepackage{url}
\usepackage{booktabs}
\usepackage{multirow}
\usepackage[hidelinks]{hyperref}

\makeatletter
\newcommand\footnoteref[1]{\protected@xdef\@thefnmark{\ref{#1}}\@footnotemark}
\makeatother

\begin{document}

\begin{acronym}
    \acro{ACE}{Acoustic Characterization of Environments\acroextra{. A noisy reverberant speech corpus and IEEE challenge run by the SAP group at Imperial College}}
\acro{AIR}{Acoustic Impulse Response}
\acro{ASR}{Automatic Speech Recognition}
\acro{CNN}[CNN]{Convolutional Neural Network}
\acro{CRNN}[CRNN]{Convolutional Recurrent Neural Network}
\acro{CTC}[CTC]{Connectionist Temporal Classification}
\acro{DNN}{Deep Neural Network}
\acro{FDDRR}[FDDRR]{Frequency-Dependent \ac{DRR}}
\acro{DRR}{Direct-to-Reverberant Ratio}
\acro{ELRR}[ELRR]{Early-To-Late Reverberation Ratio}
\acro{FDDRR}[FDDRR]{Frequency-Dependent \ac{DRR}}
\acro{FDRT}[FDRT]{Frequency-Dependent Reverberation Time}
\acro{FF}[FF]{Feed Forward}
\acro{GMM}{Gaussian Mixture Model\acroextra{. An approximation to an arbitrary probability density function that consists of a weighted sum of Gaussian distributions}}
\acro{GRU}[GRU]{Gated Recurrent Unit}
\acro{MFCC}{Mel-frequency Cepstral Coefficient}
\acro{NBC}[NBC]{Naive Bayes Classifier}
\acro{NSV}{Negative-Side Variance}
\acro{QMUL}[QMUL]{Queen Mary, University of London}
\acro{RNN}{Recurrent Neural Network}
\acro{ReLU}[ReLU]{Rectified Linear Unit}
\acro{SED}[SED]{Sound Event Detection}
\acro{STFT}{Short Time Fourier Transform}
\acro{SVM}{Support Vector Machine}
\acro{T30}[$T_\textrm{30}$]{Reverberation Time\acroextra{ to decay by $30$ dB}}
\acro{T60}[$T_\textrm{60}$]{Reverberation Time\acroextra{ to decay by $60$ dB}}
\acro{TD}[TD]{Time Distributed}
\acro{UBM}[UBM]{Universal Background Model}

\end{acronym}

\title{End-to-End Classification of Reverberant Rooms using DNNs}

\author{Constantinos~Papayiannis,~\IEEEmembership{Member,~IEEE,}
    Christine~Evers,~\IEEEmembership{Senior~Member,~IEEE,}
    and~Patrick~A.~Naylor,~\IEEEmembership{Fellow,~IEEE}
		\thanks{{\copyright}2020 IEEE.  Personal use of this material is permitted.  Permission from IEEE must be obtained for all other uses, in any current or future media, including reprinting/republishing this material for advertising or promotional purposes, creating new collective works, for resale or redistribution to servers or lists, or reuse of any copyrighted component of this work in other works.}
    \thanks{Constantinos Papayiannis is currently with Amazon Alexa, Cambridge MA, USA (e-mail:
    	papayiac@amazon.com). This work was done prior to that when he was with the Department of Electrical and Electronic Engineering, Imperial College London. Christine Evers was also with the Department of Electrical and Electronic Engineering, Imperial College London. She is now with the School of Electronics and Computer Science, University of Southampton, Southampton SO17 1BJ, U.K. (e-mail: c.evers@soton.ac.uk). P. A. Naylor is with the Department of Electrical and Electronic
Engineering, Imperial College London, London SW7 2AZ, U.K. (e-mail: p.naylor@imperial.ac.uk).}

\thanks{This work received support from the UK EPSRC Fellowship Grant EP/P001017/1, awarded to C. Evers while at Imperial College London.}}

\maketitle

\begin{abstract}
Reverberation is present in our workplaces, our homes, concert halls and theatres.   This paper investigates how deep learning can use the effect of reverberation on speech  to classify a  recording  in terms of the room in which it was recorded. Existing approaches in the literature rely on domain expertise to manually select acoustic parameters as inputs to classifiers. Estimation of these  parameters from reverberant speech is adversely affected by estimation errors,  impacting the classification accuracy. In order to overcome the limitations of  previously proposed methods, this paper shows how \acsp{DNN} can perform the classification   by operating directly on reverberant speech spectra and a   \acs{CRNN}  with an attention-mechanism  is proposed for the task. The relationship is investigated between the reverberant speech representations learned by the   \acsp{DNN} and acoustic parameters.  For evaluation, \acsp{AIR} are used from the ACE-challenge dataset that were measured in 7 real rooms. The classification accuracy of the \acs{CRNN} classifier in the experiments  is  78\% when using 5 hours of training data and 90\% when using 10 hours. 
\end{abstract}


\IEEEpeerreviewmaketitle

\section{Introduction}

Acoustic environments shape and define aspects of the sounds we hear and through this process we experience the world around us from an audible perspective. At the same time, audio recordings provide listeners with  cues that enable the understanding of properties of the environments \cite{Papayiannis2017}.  For example, as human listeners,  we are able to tell whether we are sitting in a large concert hall compared to a tiled bathroom as the two would have very different acoustics. Discriminative models can be used to allow machines to make similar distinctions between different types of acoustic environments \cite{Moore2014}. The ability to classify a recording in terms of the room in which it was recorded is becoming very important as  the adoption of smart-home devices is increasing. For instance, knowing the room in which a speaker is located    is critical  to a machine that will take action to the command "\textit{Turn on the lights!}". 

Classifying the reverberation effect has been addressed in the literature in the past. The work in \cite{Peters2012} proposed  the use of a \ac{GMM}-\ac{UBM} classifier using \acp{MFCC}. The approach in \cite{Moore2018nsv} proposed the use of the  \ac{NSV} for the classification using a  Gaussian \ac{NBC}. In \cite{Moore2014} \acp{FDRT} were used as input features to a \ac{NBC}. The work in \cite{Papayiannis2017} showed that spectral and energy-decay features, such as the \acp{FDRT} can be used for room classification and can be reliably  derived from \acp{AIR}. However, their estimation from reverberant speech  is known to be challenging \cite{Eaton2015d}.  Errors  in the estimation of acoustic parameters impact the classification accuracy of methods that use them \cite{PapayiannisThesis}.   In contrast to room classification, deep learning has been used in the past to learn other properties of reverberant environments, such as the reverberation time \cite{Cox2001,Xiong2019,gamper2018blind}, the \ac{ELRR} \cite{Xiong2019} and the room volume \cite{Genovese2019}.

The aim of this work is to overcome the limitations of  previously proposed room classification methods by proposing state-of-the-art classifiers for the task.    \acp{DNN} are used to perform the classification directly from spectrograms extracted from reverberant speech, avoiding the challenging estimation of parameters such as the \acp{FDRT}.  Working only with reverberant speech spectra also avoids time-consuming \ac{AIR} measurements  \cite{Farina2000}. The performance of a set of candidate \ac{DNN} architectures is evaluated. The best performing model is a \ac{CRNN} that incorporates an attention-mechanism.  \acp{CRNN} have been successfully used for \ac{ASR} \cite{Sainath2017},  \ac{SED} \cite{Yan2019}, and media-presence detection \cite{Papayiannis2018}. Attention-mechanisms have provided substantial  performance improvements for \ac{ASR} \cite{Chan2019las}, translation \cite{BahdanauCB14} and  image-description \cite{XuBKCCSZB15}. Motivated by their success in other areas, \acp{CRNN} and attention-mechanisms are now applied to the task of room classification. The proposed combination shows benefits compared to the other approaches considered.  The performance of the proposed model is compared to that  of previous classifiers in the literature.  Important  contributions of this work  include the use of deep learning for room classification and the insight derived from the analysis of the attention-vectors and representations learned by \acp{DNN}. This  reveals what  information is identified as important to the task during training and the relationship of this information with acoustic parameters.

The structure of the remainder of this paper is as follows: Section \ref{section_chapter3a_signal_model} presents the  model of reverberant speech. Section \ref{chapter3a_section_discriminative_nets_intro} discusses the candidate \ac{DNN} architectures for the task. Section \ref{section_training} presents the method used to train \acp{DNN} and the data used. Section \ref{section_c3a_experiments} presents the experiments that evaluate the performance of  \acp{DNN} on the task of room classification. Section \ref{section_representation} analyzes the representations learned by the  \acp{DNN} and their relationship with acoustic parameters. Conclusions are drawn in Section \ref{section_c3a_conclussion}.

\section{Signal model}
\label{section_chapter3a_signal_model}

In a dataset of size $M$, the $m^\text{th}$ reverberant speech signal ${\mathbf{x}_m \triangleq [x_m(0),\dots,x_m(N-1)]}$,  is defined as 
\begin{equation}
	x_m(n) = h_m(n) \ast s_m(n),
\end{equation}
where $\ast$ describes the convolution operation,  ${\mathbf{s}_m \triangleq [s_m(0),\dots,s_m(N-1)]}$ the anechoic speech signal for sample index  $ n \in \{0,\dots,N-1\}$ and ${\mathbf{h}_m \triangleq [h_m(0),\dots,h_m(N_h-1)]}$  denotes the \ac{AIR} of length $N_h$.

Vectors $\mathbf{x}_m$, $\mathbf{s}_m$ and $\mathbf{h}_m$ are transformed into the log-power \ac{STFT} domain to obtain $\mathbf{X}_m$, $\mathbf{S}_m$ and $\mathbf{H}_m$ respectively, similar to \cite{Papayiannis2018}. The size of the \ac{STFT} frames $N_s$  is  selected for relevant experiments in later Sections. An overlap of $\frac{N_s}{2}$ samples exists between each \ac{STFT} frame.

\section{Candidate \acs{DNN} architectures}
 \label{chapter3a_section_discriminative_nets_intro}

\begin{figure*}[t]
	\centering
	\subfloat[\acs{CNN}]{
		\includegraphics{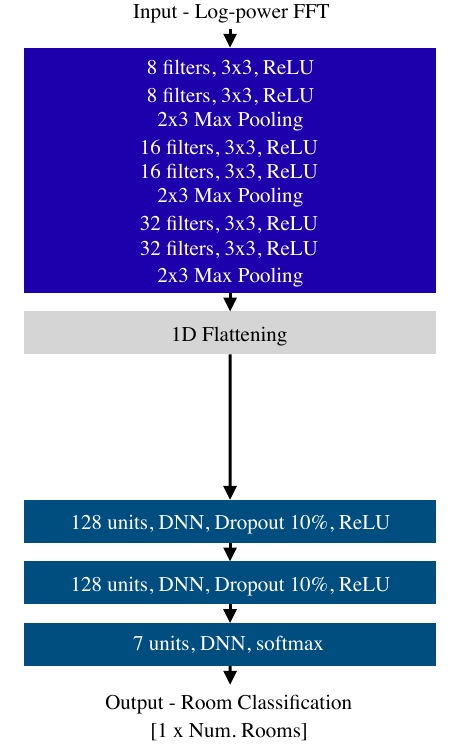}
		\label{figure_Chapter3a_models_speech_cnn}
	}~
	\subfloat[\acs{RNN}]{
		\includegraphics{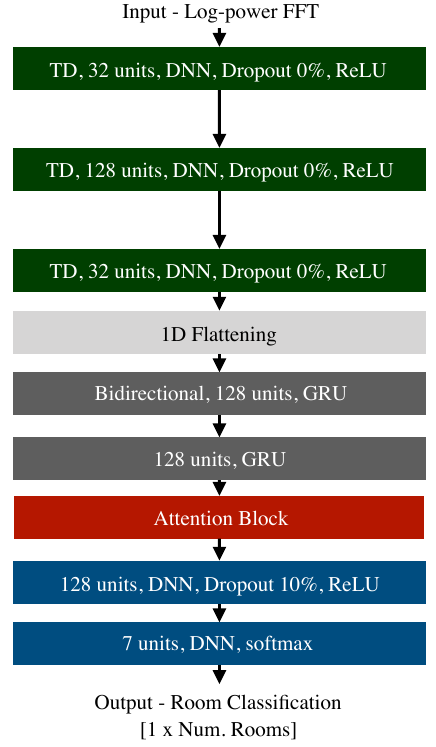}
		\label{figure_Chapter3a_models_speech_rnn}
	}~
	\subfloat[\acs{CRNN}]{
		\includegraphics{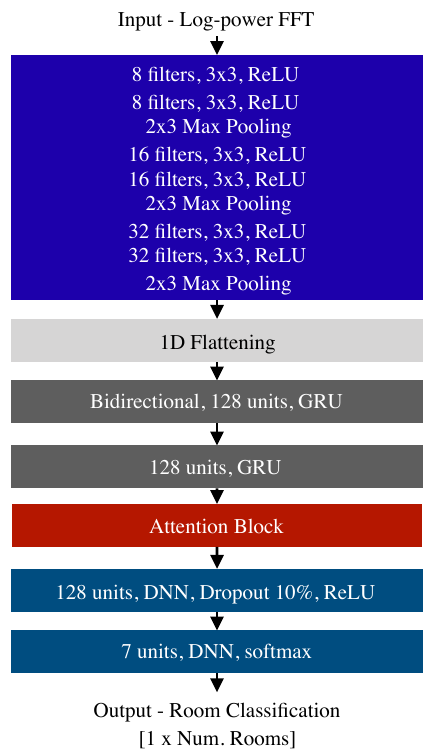} 
		\label{figure_Chapter3a_models_speech_cnn_rnn}
	}
	\caption{Candidate architectures for room classification from reverberant speech.}
	\label{figure_Chapter3a_models_speech}
\end{figure*}

 \subsection{Architectures' Description}

The first candidate model is a   \acf{CNN}. \acp{CNN} have been previously used for  \ac{ASR} using audio spectrograms \cite{cnn_asr2014} and for  \ac{T60} estimation \cite{gamper2018blind}. Inputs to the \acp{CNN} are processed by a stack of convolutional layers  \cite{Simonyan2014} that apply small filters  to the input data. Convolutional  layers are typically separated by Max Pooling layers \cite{Zhou1988} to reduce  dimensionality while maintaining the most relevant information.  Dropout \cite{Hinton2012} is used at the input of fully-connected layers to avoid overfitting. The \ac{ReLU}  function \cite{Hahnloser2000} is typically used as an activation. A softmax activation is applied to the output of the last fully connected layer.
 To incorporate the sequential nature of speech signals, the second candidate model is a \ac{RNN}. The inputs to the \acp{RNN} are first processed by \ac{TD} layers. A \ac{TD} layer with  output  $\mathbf{Y} \in \mathbb{R}^{N_f \times D_{y} }$ performs the operation
	\begin{equation}
	\mathbf{y}_i = \mathbf{W} \mathbf{x}_{i} + \mathbf{b}~\forall~ i \in \{0,\dots,N_f-1\},
\end{equation}
where $\mathbf{y}_i$ are rows of  $\mathbf{Y}$ and $\mathbf{x}_i $ are rows of the input $\mathbf{X} \in \mathbb{R}^{ N_f \times D_x  }$. $\mathbf{W}  \in \mathbb{R}^{ D_y \times D_x }$ and $\mathbf{b}  \in \mathbb{R}^{ D_y }$ are learnable parameters. 
The 3$^{\text{rd}}$ candidate model combines convolutional and recurrent layers to form a \ac{CRNN}  \cite{Sainath2015}. All the architectures are given in   Fig.~\ref{figure_Chapter3a_models_speech}.  The \ac{RNN} and \ac{CRNN} networks are studied with and without an attention-mechanism, which is described in the following section.

 \subsection{Attention  Mechanism}
 \label{section_attention}

The \ac{RNN} and \ac{CRNN} architectures for room classification shown in Fig.~\ref{figure_Chapter3a_models_speech} incorporate an attention-mechanism. The mechanism is used to  allow each model to  emphasize or de-emphasize the importance of individual time-frames. The mechanism ``guides'' the classifier to make more accurate predictions by treating \ac{RNN} layers as encoders and computing an attention-vector ${\alpha_i}$  for their output sequence ${\zeta_i}$, where $i \in \{0,...,N_i-1\}$ a time-step of the recurrent layer and $N_i$ the total number of  time-steps. 

The attention-mechanism is implemented as \cite{BahdanauCB14,XuBKCCSZB15}
\begin{equation}
\alpha_i = \frac{\exp{\{a_{f}(\zeta_i)\}}}{\sum_{i=0}^{N_i-1}\exp\{a_{f}(\zeta_i)\}},
\label{eq_attention_alpha}
\end{equation}
where $\alpha_i$ the attention applied to frame $i$ and $a_{f}$ a function parametrised by a 1-layer feed-forward network. The mechanism operates on the sequence of the last recurrent layer and the resulting context vector 
\begin{equation}
	\mathbf{c} = \sum_{i=0}^{N_i-1} \alpha_i  \zeta_i
\end{equation}
is used as the input to  fully-connected layers, as shown in Figs.~\ref{figure_Chapter3a_models_speech_rnn} and  \ref{figure_Chapter3a_models_speech_cnn_rnn}.

\section{Network training and generalization}
\label{section_training}

\subsection{Dataset Generation}
\label{section_dataset}

The candidate models are trained using spectrograms of reverberant speech signals, obtained by convolving anechoic speech signals from TIMIT \cite{Garofolo1993} with \acp{AIR} from the \ac{ACE} challenge \cite{Eaton2015a}  and  \ac{QMUL} datasets \cite{Stewart2010}, downsampled to 16~kHz. The \ac{ACE} dataset contains  \acp{AIR} measured in 7 rooms  (see Table \ref{table_room_info}) using 5 microphone-arrays. The arrays contain between 2 to 32 microphones. 100 \acp{AIR}  are provided for each room at 10 different microphone-array positions, giving a total of 700 \acp{AIR} at 70 array positions.  Using the  \ac{QMUL} dataset allows the evaluation of the generalisation of the presented training methods and model architectures. \ac{ACE} \acp{AIR} were predominantly recorded in small offices and meeting rooms whereas  \ac{QMUL} \acp{AIR} were recorded in much larger environments such as halls (see Table \ref{table_room_info}). The \acf{T60} and \ac{DRR} for each \ac{AIR} from each room in  \cite{Eaton2015a} and \cite{Stewart2010} are shown in Fig.~\ref{figure_rooms_t60_drr}. 

Anechoic speech utterances are created by concatenating sentences from a single speaker from the  TIMIT dataset. Each resulting utterance is  convolved with one \ac{AIR} to produce reverberant speech. Each \ac{AIR} is convolved with $N_u$ speech utterances. Prior to using a speech utterance, it is randomly  time-shifted and truncated to 5~s. The selection of which utterance to convolve with an \ac{AIR} is random with replacement, which means that the speech data are oversampled.  It is important to note that speech from the same speaker is never used for both training and testing and the training and test splits of the  TIMIT distribution are maintained. The resulting reverberant speech samples are used to create a matrix $\mathbf{X}$ of dimensions $\left[ N_u \times M , N_f, \frac{N_s}{2}+1 \right]$. The framesize $N_s$  for the analysis of speech is set to 20~ms (320 samples) and a  Hanning window is applied to each frame. For the utterance length of 5~s and a sampling rate of 16~kHz, $N_f=500$. Values for the number of utterances convolved per \ac{AIR}, $N_u$, and the total number of \acp{AIR},  $M$, are varied for each experiment as detailed in Section \ref{section_evaluation}.

\subsection{Training data batches}

\begin{figure}[t]
	\centering       	
	\includegraphics[scale=0.65]{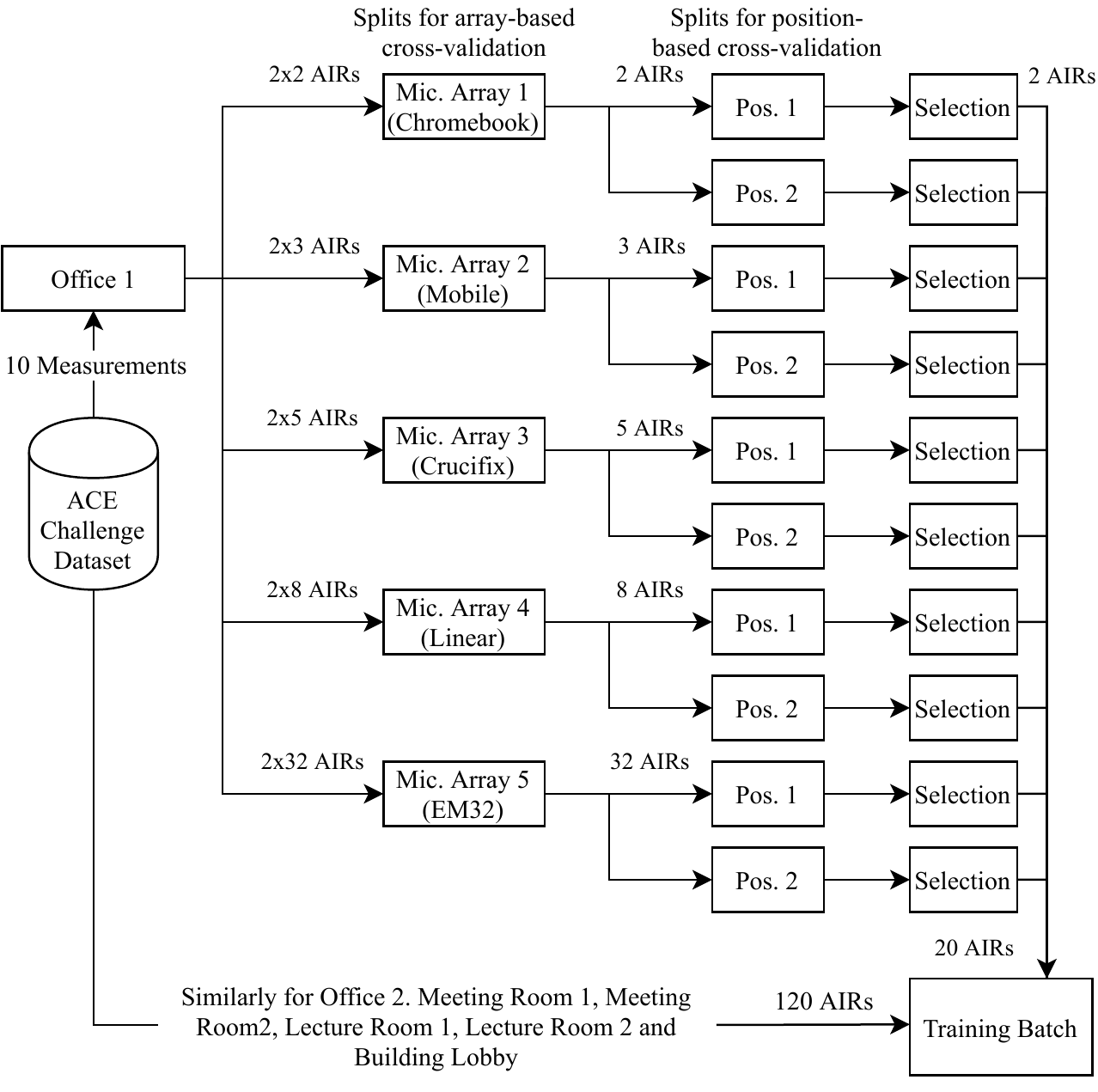}

	\caption{Organization of the data from the \ac{ACE} dataset according to rooms, microphone-arrays and recording positions for the purpose of training \ac{DNN} room classifiers. The ``Selection'' block maintains a counter looping through the \acp{AIR} contained in the previous block. }
	\label{figure_ace_batches}
\end{figure}

\label{section_batch_contruction}

\label{section_c3a_batch_balancing}
The  \ac{ACE} \ac{AIR} data \cite{Eaton2015a} is evenly distributed across rooms, however, it is significantly imbalanced in terms of the number of \acp{AIR} available for each measurement position. This proves problematic when constructing training batches  by randomly sampling the dataset as performance is biased towards the modes that contributed most of the training data. Therefore, batches are constructed  with a   balanced number of samples from each of the measurement positions. The method   includes in each batch a fixed and equal number of \acp{AIR} from each of the measurement positions. This  creates a balance across the positions and, for an equal number of positions per microphone-array and the same set of microphone-arrays used per room, this simultaneously creates a balance across rooms.  The batch construction method is shown in Fig.~\ref{figure_ace_batches}. The ``Selection'' block is a  mechanism that iterates  through the \acp{AIR} contained in the previous block. Each block selects and contributes two  \acp{AIR} to each batch, which are convolved with anechoic speech. For the case of the \ac{ACE} challenge shown in Fig.~\ref{figure_ace_batches}, the resulting batch size is $M_b=140$. 

During training, an epoch is defined by $\lceil M N_u \rceil$ weight updates, with one batch constructed for each. In the method described above for creating batches, \acp{AIR}  are reused within the same epoch.  
As stated in Section \ref{section_dataset}, for each \ac{AIR} and each batch, reverberant speech is created using randomly selected and shifted samples from the  TIMIT dataset.  These  random permutations of the data avoid the repeated presentation of identical training samples to the model during training and the subsequent bias and overfitting to specific examples. 

\subsection{Optimization} 

The Adam optimizer is used to train the networks by minimizing the categorical cross-entropy between the model's output and the label for the room associated with each sample.  To avoid overfitting, early stopping is used  to terminate the training when  the validation loss has not improved for 10 epochs. Models are trained for  a maximum of  50 epochs. The validation set is constructed by  stratified selection across rooms and amounts to 15\% of the available data for training. 

\label{section_c3a_early_stopping}

\begin{table}[t!]
	\centering
	\begin{tabular}{|r|r|r|r|r|r|}
		\hline
		Dataset & Room & \multicolumn{1}{c|}{\begin{tabular}[c]{@{}c@{}}L\\ (m)\end{tabular}} & \multicolumn{1}{c|}{\begin{tabular}[c]{@{}c@{}}W\\ (m)\end{tabular}} & \multicolumn{1}{c|}{\begin{tabular}[c]{@{}c@{}}H\\ (m)\end{tabular}} & \multicolumn{1}{c|}{\begin{tabular}[c]{@{}c@{}}Vol.\\ (m$^3$)\end{tabular}} \\ \hline
		\multirow{7}{*}{ACE \cite{Eaton2015a}} & Office 1 & 3.32 & 4.83 & 2.95 & 47.30 \\
		& Office 2 					 	 & 3.22 & 5.10 & 2.94 & 48.30 \\ 
		& Meeting Room 1 		 & 6.61 & 5.11 & 2.95 & 99.60 \\ 
		& Meeting Room 2 		& 10.30 & 9.07 & 2.63 &246.00\\
		& Lecture Room 1 		 & 6.93 & 9.73 & 3.00 &202.00\\ 
		& Lecture Room 2 		& 13.60 & 9.29 & 2.94 & 370.00 \\ 
		& Building Lobby           & 4.47 & 5.13 & 3.18 & 72.90  \\ \hline
		\multirow{3}{*}{QMUL \cite{Stewart2010}} & Classroom & 7.50 & 9.00 & 3.50 & 236.00 \\ 
		& Great Hall$^*$ & 16.00 & 23.00 & \multicolumn{1}{r|}{$^{**}$} & \multicolumn{1}{r|}{$^{**}$} \\ 
		& Octagon$^*$ & 23.00 & 23.00 & 21.00 & 9500.00 \\ \hline
	\end{tabular}
\begin{flushright}
 \footnotesize{$^*$Room is not ``shoe-box'' shaped.},	
	\footnotesize{$^{**}$Information not provided by authors.}
\end{flushright}
	\caption{Datasets of \acp{AIR} used in this work and the corresponding room information. }
	\label{table_room_info}
\end{table}
\begin{figure}[t!]
	\centering	
	\includegraphics[scale=0.7]{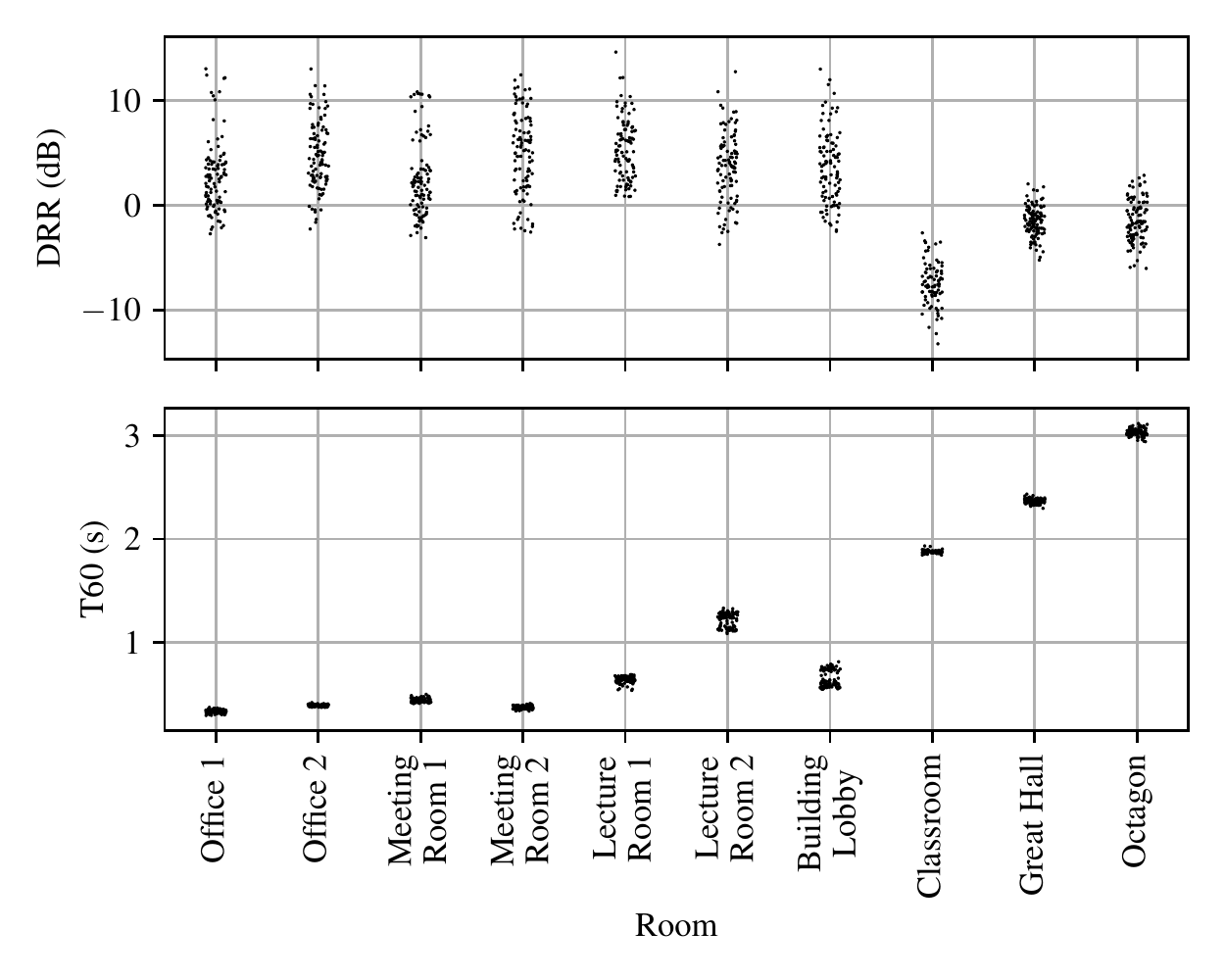}	
	\caption{\ac{DRR} and \ac{T60} values for the \acp{AIR} used in this work grouped by room.}
	\label{figure_rooms_t60_drr}

\end{figure}

\subsection{Evaluation}
\label{section_evaluation}

Cross-validation is used for the evaluation of classifiers. For the \ac{ACE} dataset, two types of partitioning of the dataset into folds are studied. The first type, similar to \cite{Papayiannis2017}, partitions the dataset into 70 folds,  each one corresponding to a recording position (see Fig.~\ref{figure_ace_batches}). The second  partitions the data into 5 folds, each one corresponding to the microphone-array used for each measurement. The second case is  more challenging as more data is held-out  during training and the microphone-array used for recording the held-out samples is unseen during training.  For the \ac{QMUL} dataset the data is split based on positions on the densely populated  two-dimensional measurement grid \cite{Stewart2010}. In order to ensure dissimilarity of data points between the splits, the receiver-position  grid is split in three uniformly sized parts along the x-axis value of the microphone's position. The two extreme ends are kept and each one forms one fold. In total, 288 \acp{AIR} are used from the  dataset. Two uniformly sized folds are therefore used to split 288 \acp{AIR} from the three rooms that form the \ac{QMUL} dataset. Each fold is tested by predicting the class of the held-out samples by training a network using the remaining samples. As the evaluation measure, the accuracy of predictions is used. 

The baselines considered for the task are the following: Baseline~1 is the method proposed in \cite{Moore2014}, which is based on the use of \acp{FDRT} as ``\textit{Roomprints}'', combined with Gaussian \acfp{NBC}. Baseline~2 uses the same inputs with a 3-layer \ac{FF} network as the classifier with 32 units for each hidden layer and  dropout of probability of 10\%. This is equivalent to replacing the \ac{NBC} classifier of \cite{Moore2014} with a \ac{DNN} classifier. Baseline~3 is the \acs{SVM} classifier and features proposed in \cite{Papayiannis2017}. The features are a collection of acoustic parameters that include \acp{FDRT}, \acp{MFCC} and \acp{FDDRR}.  All of the above baselines take  acoustic parameters as inputs, which are estimated with access to the \acp{AIR}. The performance of the \acp{DNN} is also compared with the method of \cite{Peters2012}. It is based on a \ac{GMM}-\ac{UBM} setup, training a base \ac{GMM} model with 128 mixtures, that is later copied and adapted to form one \ac{GMM} for each of the rooms. The inputs to the model are reverberant speech \acp{MFCC}, augmented with $\Delta$s and $\Delta\Delta$s of the parameters. This is an alternative to the use of \acp{DNN}.

\section{Experimental Evaluation}

\label{section_c3a_experiments}

\begin{table*}[t]
	\centering
	\begin{tabular}{|r|c|ccc|c|ccccc|}
		\hline
		\multicolumn{1}{|c|}{\multirow{2}{*}{\begin{tabular}[c]{@{}c@{}}Datasets\\ included\end{tabular}}} & \multirow{2}{*}{\begin{tabular}[c]{@{}c@{}}ACE folds\\ cross-validation\end{tabular}} & \multicolumn{3}{c|}{AIRs} & \multicolumn{6}{c|}{Reverberant Speech} \\ \cline{3-11} 
		\multicolumn{1}{|c|}{} &  & \begin{tabular}[c]{@{}c@{}}Baseline 1 \cite{Moore2014}\\ FDET NBC\end{tabular} & \begin{tabular}[c]{@{}c@{}}Baseline 2\\ FDRT FF\end{tabular} & \begin{tabular}[c]{@{}c@{}}Baseline 3 \cite{Papayiannis2017}\\ SVM Mult.\end{tabular} & \begin{tabular}[c]{@{}c@{}}Baseline 4 \cite{Peters2012}\\ GMM-UBM\end{tabular} & RNN & Att.-RNN & CNN & CRNN & Att.-CRNN \\ \hline
		QMUL & -- & \textbf{1.000} & 0.851 & \textbf{1.000} & 0.610 & 0.894 & \textbf{0.936} & 0.896 & 0.852 & 0.899 \\ \hline
		ACE & Position & 0.981 & 0.966 & \textbf{1.000} & --$^*$ & 0.793 & 0.770 & 0.732 & 0.865 & \textbf{0.896} \\ \hline
		ACE & Mic. Array & 0.568 & 0.752 & \textbf{0.874 }& 0.693 & 0.730 & 0.798 & 0.714 & 0.777 & \textbf{0.904} \\ \hline
		ACE \& QMUL & Mic. Array & 0.680 & 0.769 & \textbf{0.844} & 0.620 & 0.718 & 0.814 & 0.690 & \textbf{0.821} & 0.787 \\ \hline
	\end{tabular}
	\begin{flushright}
	$^*$ Not evaluated due to large number of folds and long training times required by the baseline.
\end{flushright}
	\caption{Results of room classification as the accuracy of each classifier. Baselines 1--3 rely on the availability of \acp{AIR} whereas Baseline 4 uses reverberant speech \acp{MFCC}. \ac{DNN} classifiers operate directly on reverberant speech spectra. }

		\label{table_chapter3a_accuracy_airs}
\end{table*}

\label{section_air_class}

\begin{figure}[t]
	\centering       	
	\includegraphics[scale=1.]{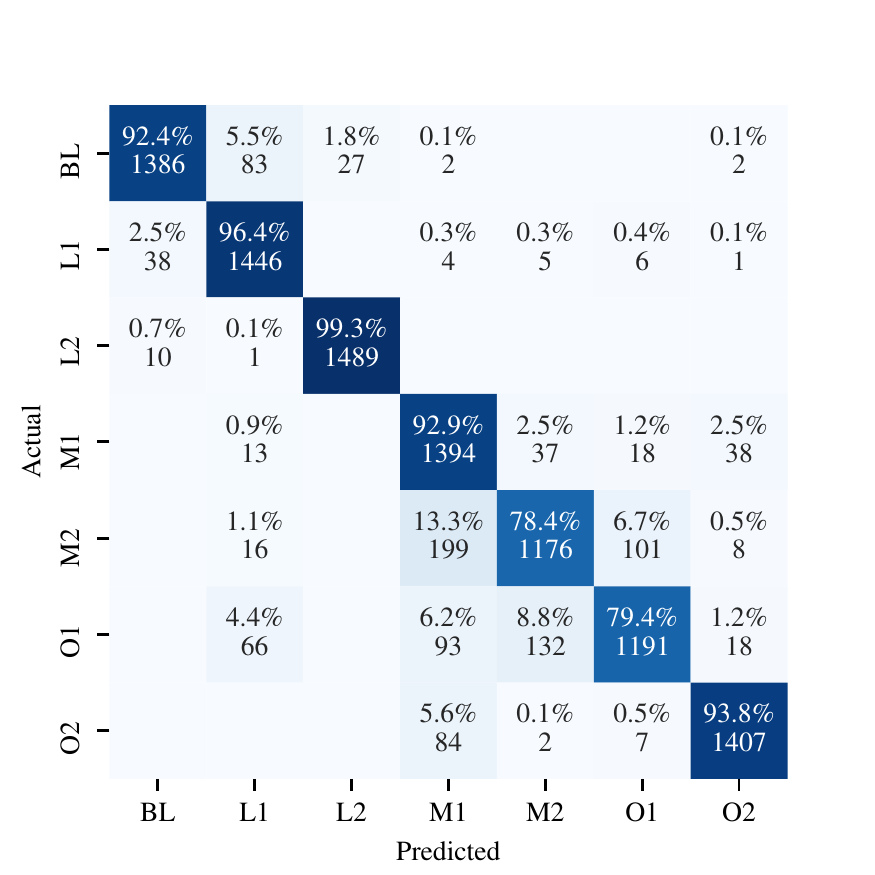}
	
	\small{Legend: MR1~-~Meeting~Room~1, MR2~-~Meeting~Room~2,  BL~-~Building~Lobby, O1~-~Office~1, O2~-~Office~2, LR1~-~Lecture~Room~1, LR2~-~Lecture~Room~2.}
	\caption{Confusion matrix for room classification based on reverberant speech by the Attention-\ac{CRNN} classifier using data from the \acs{ACE} and TIMIT datasets.  }
	\label{figure_Chapter3a_speech_confusion}
\end{figure}

The classification accuracy of the \acp{DNN}   and the baselines  described in Section \ref{section_evaluation} are summarized in Table \ref{table_chapter3a_accuracy_airs}.  The results show that for the \ac{ACE} challenge dataset,  the best performing \ac{DNN} architecture for the task is consistently the Attention-\ac{CRNN}, which achieves accuracies of 90\% for both cross-validation types.  The superior performance of the specific architecture is attributed to the fact that it features the most diverse set of layers, first using convolutional layers to extract feature-maps of reverberant speech, which are then processed as a sequence by bidirectional recurrent layers before the attention mechanism is applied. The confusion matrix associated with the microphone-array based cross-validation is shown in Figure \ref{figure_Chapter3a_speech_confusion}.  For the case of the \ac{QMUL} dataset, the task is much simpler as 3 rooms are involved, compared to the case of the \ac{ACE} challenge, which involves 7 rooms. All classifiers perform better in this case. The  best accuracy is provided by the Attention-\ac{RNN} at 94\%.  The Attention-\ac{CRNN} achieves an accuracy of 90\%. While its accuracy is almost identical to that on the  \ac{ACE}  dataset, the remaining models show significant differences between the two cases. The generalization of the Attention-\ac{CRNN} to both datasets is illustrated by this observation. When considering the joint case of \ac{ACE} and \ac{QMUL}, the best performing model is the \ac{CRNN} model, with an accuracy of 82\%. The Attention-\ac{RNN} is a close second with an accuracy of 81\%. The  Attention-\ac{CRNN} gives a comparable accuracy of 79\%. The remaining models show significantly lower performance on this most challenging test-case. 

Comparing the performance of the \acp{DNN} with Baseline 4 shows that all architectures outperform the baseline in all cases. The remaining  baselines assume access to the \acp{AIR} to estimate acoustic parameters as features. This makes the task significantly easier as the acoustic channel does not have to be inferred from reverberant speech. Estimation of such parameters from reverberant speech is a significantly challenging task \cite{Eaton2015d} and estimation errors impact the performance of the classifiers \cite{PapayiannisThesis}. Nevertheless, the \ac{DNN} classifiers provide comparable or even greater accuracy than these baselines in Table \ref{table_chapter3a_accuracy_airs}. For the case of the \ac{ACE} challenge dataset and with microphone-arrays as the cross-validation splits, the Attention-\ac{CRNN} provides an accuracy of 90\%, compared to 84\% provided by \cite{Papayiannis2017}. The method of \cite{Papayiannis2017} is consistently the best performing baseline. However, it uses \ac{FDDRR} values, which are notoriously difficult to estimate without access to \acp{AIR}  \cite{Eaton2015d}. The easiest task for the baselines is the case of the \ac{QMUL} dataset. This is attributed to the well separated \ac{T60} values of the \acp{AIR} in the dataset (see Fig.~\ref{figure_rooms_t60_drr}). As all classifiers use \acp{FDRT} derived from \acp{AIR}, the task becomes trivial. The same does not apply for the case of the \ac{ACE} challenge, which is reflected in the accuracy numbers. Notably, classifying \acp{AIR} with the  microphone-arrays as splits makes the task much more challenging over the case of having only the positions as splits, both for \acp{DNN} and the baselines. For baselines with \acp{FDRT} as the features this is particularly the case with the accuracy of  \cite{Moore2014} falling to 57\% from 98\%. Using a fully-connected layer instead of the \ac{NBC} provides respective scores of 97\% and 75\%. This indicates that information about the microphone-array used during inference is important for \ac{FDRT} based room classifiers.

\begin{figure}[t]
	\centering       	
	\includegraphics[scale=0.8]{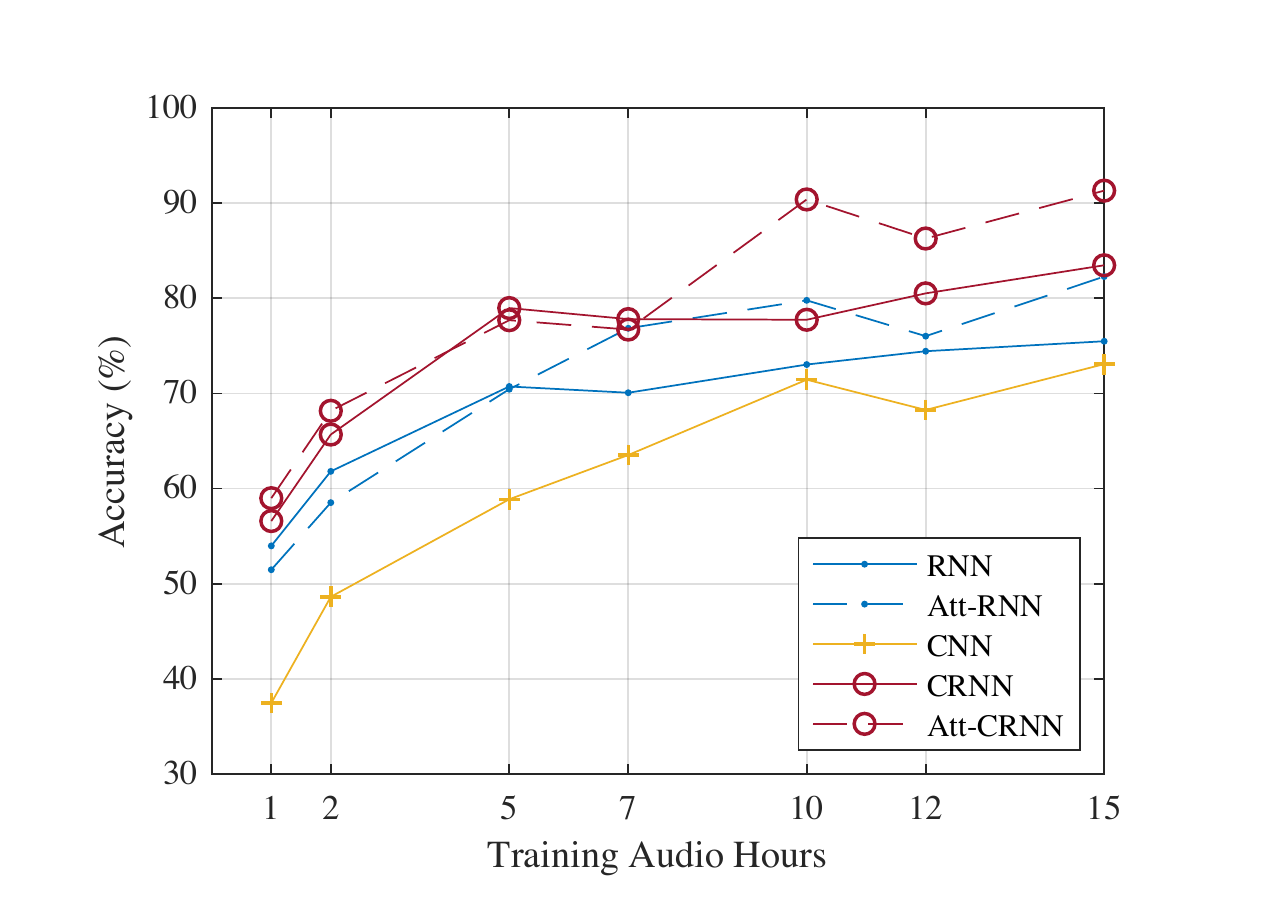}
	
	\caption{Accuracy of \ac{DNN} classifiers versus available training audio hours for the \ac{ACE} dataset. The test set is fixed to 15 hours of reverberant speech, equally distributed across rooms.}
	\label{accuracy_vs_hours}
\end{figure}

Considering the potential applications of the classifiers, an important dimension to consider is the amount of audio that is needed to train them at a satisfactory level of accuracy. Such an analysis for each of the architectures studied is performed and the results are shown in Fig.~\ref{accuracy_vs_hours}. The figure shows the accuracy of each network when trained with an increasing amount of training hours. All models trained are tested on a static test set of 15~hours. It can be seen that with less than 2 hours of available training hours the performance of the networks is significantly worse than having 15~hours available. The most significantly impacted is the \ac{CNN} model that shows a drop of 49\% relative accuracy from the highest of 73\%. With five hours available, all networks achieve more than 80\% of relative performance compared to having 15 hours of audio available for  training. The Attention-\ac{CRNN} results in an accuracy of 78\%.  It is also interesting to note that, while for other networks providing more than 7 hours of training data offered little improvements, the Attention-\ac{CRNN} continued to improve beyond that point, reaching an accuracy of 91\% with 15~hours used for training. This behaviour is attributed to the more diverse set of mechanisms that are part of the model, allowing it to more effectively leverage spectral and temporal patterns in the input, which are enhanced by the ability to learn to attend to the most informative parts of its input. As shown in (\ref{eq_attention_alpha}), information from all time-steps are weighted and effectively used for the classification, compared to the case of not using the attention mechanism that forces the recurrent layer to retain all important information in the last time-frame, which is driving the final classification. The attention-mechanism also offers a  benefit to the \ac{RNN} model but with an overall lower accuracy than the \ac{CRNN} equivalent.

\section{Analysis of learned representations}
\label{section_representation}

\begin{figure}[t]
	\centering
	\includegraphics[scale=0.6]{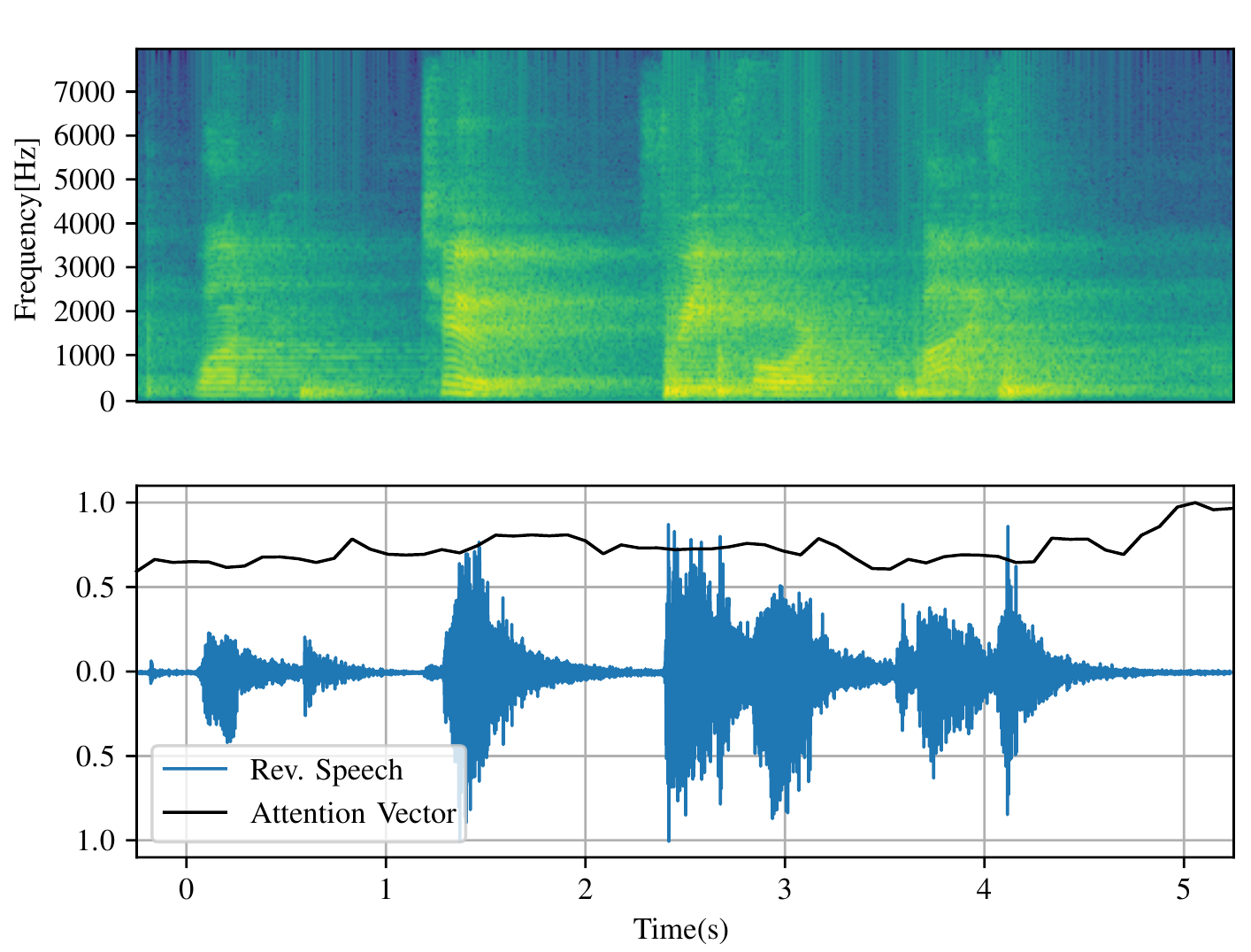}
	\caption{Reverberant speech spectrogram, waveform and attention vector applied by the Attention-\ac{CRNN}.}
	\label{figure_attention_speech}
\end{figure}

\begin{table}[t]
	\centering
	\begin{tabular}{@{}rccc@{}}
		\toprule
		& CNN    & CRNN  & Rand. Init. \\ \midrule
		DRR $\rho_{ccc}$     & 0.257  & 0.260 & -0.039      \\
		T30 $\rho_{ccc}$    & 0.038  & 0.022 & 0.036       \\
		T60 $\rho_{ccc}$    & -0.004 & 0.046 & -0.030      \\ \bottomrule
	\end{tabular}
	\caption{Accuracy of prediction of acoustic parameters using   \ac{DNN} embeddings. Accuracy is measured by the concordance correlation coefficient $\rho_{ccc}$ between the predicted and estimated values of the parameters from the \ac{AIR}. }
	\label{corrs_cnn}
	
	\begin{tabular}{@{}rcc@{}}
		\toprule
		Spectral Feature & Attention-CRNN  & Rand. Init. \\ \midrule
				Centroid  $z_c(i)$ & -0.115 & 0.002 \\
		Bandwidth $z_b(i)$   & -0.188 & -0.005 \\
		Roll-off   $z_r(i)$  & -0.004 & 0.007      \\ \bottomrule
	\end{tabular}
	\caption{Pearson correlation coefficient between attention-vectors applied to reverberant speech by the Attention-\ac{CRNN} and spectral features. A randomly initialized and untrained network of identical architecture is used a reference. }
	\label{corrs_attention}
	
\end{table}

As the accuracy of different \acp{DNN} and baselines differs across datasets, an additional experiment is performed that provides insight into the relationship between acoustic parameters, namely the \ac{DRR} and \ac{T60}, and the representation of reverberant speech at the outputs of deeper layers of the networks. This experiment studies the representations at the output of the last convolutional stage of the \ac{CNN} and \ac{CRNN} of Fig.~\ref{figure_Chapter3a_models_speech}. The representation is formed by a set of output channels, each one being a higher level and lower dimensional representation of the reverberant speech spectrogram. To investigate the correlation between acoustic parameters and the representations, the values of the \acs{T30}, \ac{T60} and \ac{DRR} are linearly predicted directly from the representations. To this end, a weight vector $\mathbf{w} \in \mathbb{R}^{D_{c}}$ and a scalar $b$ are trained to minimize
\begin{equation}
l =  ( y_m - \hat{y}_m )^2
\end{equation}
where $y_m$ the value of the acoustic parameter being estimated, extracted from the \ac{AIR} $\mathbf{h}_m$, and $\hat{y}_m$ its estimate as 
\begin{equation}
\hat{y}_m = \mathbf{w} f_{\text{conv}}(\mathbf{X}_m) + b.
\end{equation}
The function $f_{\text{conv}}$ corresponds to the convolutional layers  of the models in Figs.~ \ref{figure_Chapter3a_models_speech_cnn} and \ref{figure_Chapter3a_models_speech_cnn_rnn}, which outputs a representation-vector of length $D_c$. The matrix $\mathbf{X}_m$ is the reverberant speech spectrogram. The values of $\mathbf{w} $ and  $b$ are learned using the Adam optimizer \cite{Kingma2014} using batches of 16 \acp{AIR}. The accuracy of the predictions is measured by the concordance correlation coefficient \cite{lin_ccc} defined as 
\begin{equation}
\rho_{ccc} = \frac{2 \rho_{y\hat{{y}}} \sigma_{{y}} \sigma_{\hat{{y}}}}{ \sigma_{{y}}^2 +  \sigma_{{\hat{y}}} ^ 2  + (\mu_{{y}}-\mu_{{\hat{y}}})^2}
\end{equation}
where $\rho_{y\hat{{y}}}$ is the Pearson's correlation coefficient between the estimates and values of parameters,  $\sigma_{{y}}$ and  $\sigma_{\hat{{y}}}$  the standard deviation for each and similarly for the means $\mu_{{y}}$ and $\mu_{\hat{{y}}}$. The results of this experiment are shown in Table \ref{corrs_cnn} and show that there is a  non-negligible correlation  $\rho_{ccc}$ of 0.3  between linear estimates of \ac{DRR} from the representations and the values of the \ac{DRR} measured from the \acp{AIR}. 

A similar analysis is performed on the attention-vectors calculated by the Attention-\acp{CRNN}. The correlation between the  values of the vectors defined in (\ref{eq_attention_alpha}) and spectral features is measured. The attention values $\alpha_i$, where $i \in \{0,\dots,N_f-1\}$ is the frame index, are calculated for a reverberant speech signal $\mathbf{x}_m$ with \ac{STFT} matrix  $\mathbf{F} \in \mathbb{R}^{ N_f \times K  }$. The correlation is evaluated between the  attention-vector and the following spectral features:
\begin{enumerate}
	\item [1.] The spectral centroid  \cite{spectral_bandwidth}, defined as 
\end{enumerate}
\begin{equation}
	z_c(i)= \frac{\sum_{k=0}^{K-1} F_k(i)  f_k}{\sum_{k=0}^{K-1} F_k(i)},
	\label{eq_spectral_centroid}
\end{equation}
\begin{enumerate}
	\item [] where $f_k$ the centre frequency of \ac{STFT} bin with index $k$ and $F_k(i)$ the magnitude of the \ac{STFT}  at frame  $i$. 
	\item [2.]The spectral bandwidth  \cite{spectral_bandwidth}, defined as 
\end{enumerate}
\begin{equation}
		z_b(i)= \left\{\sum_{k=0}^{K-1}  F_k(i)  \{f_k - z_c(i)\}^2\right\}^\frac{1}{2},
\end{equation}
\begin{enumerate}
	\item [] which is a measure of the energy away from the centroid defined by (\ref{eq_spectral_centroid}).
	\item [3.] The spectral roll-off $z_r(i)$ \cite{spectral_rolloff}, which indicates the frequency  $f_k$ that corresponds to the bin under (and including) which 85\% of the energy of  frame $i$ is captured.
\end{enumerate}
The three features are complimentary, in the sense that the centroid indicates the frequency region where energy is most concentrated, the bandwidth measures the spread of the energy across the spectrum and the roll-off indicates the frequency above which little energy exists. The results in Table \ref{corrs_attention} show the Pearson correlation coefficient between the attention-vector and each feature. The results show that all correlations are negative. The strongest negative correlation of -0.19 exists between the bandwidth of the reverberant speech signal and the attention-vector, indicating that more attention is given to frames that contain energy concentrated closer to the centroid $z_c(i)$. Smaller correlation exists between the location of the centroid and the attention-vector and no correlation is shown between the roll-off frequency $z_r(i)$ and the attention-vector. As an example, Fig.~\ref{figure_attention_speech} shows a reverberant speech signal and its spectrum, along with the corresponding attention-vector. What is characteristic in the figure is that the attention vector has higher values near the end of the utterance where energy is decaying, indicating the importance of this region for the task.
 
\section{Discussion and conclusion}
\label{section_c3a_conclussion}

This paper proposed the use of neural networks for room classification. Five model architectures were investigated as classifiers for the task: 1)~\ac{CNN} 2)~\ac{RNN} 3)~Attention-\ac{RNN} 4)~\ac{CRNN} 5)~Attention-\ac{CRNN}. The performance of the trained classifiers was compared to the \ac{GMM}-\ac{UBM} of \cite{Peters2012} using \acp{MFCC}. In the experiments presented, the Attention-\ac{CRNN} architecture provided the highest classification accuracy in most of the test cases. The model also provided the highest accuracy when very limited amounts of training hours were presented during training. Since the models are trained using spectrograms of reverberant speech, estimation errors due to acoustic parameter estimation are avoided. The experiments presented on the \ac{ACE} challenge  dataset have shown that  the Attention-\ac{CRNN} achieves a classification accuracy of 78\% using 5 hours and 90\% using 10 hours of training data.  

Convolutional layers in the networks learn representations of reverberant speech spectrograms. An investigation of the predictive power of the representations regarding a set of acoustic parameters indicated that the \ac{DRR} can be linearly predicted from the representation with a concordance correlation coefficient of 0.3 between the \ac{DRR} values and their estimates. Furthermore, a correlation of -0.19 was identified between the attention-vectors produced by the Attention-\ac{CRNN} and the spectral bandwidth of the input reverberant speech. A notable finding of the above analysis is the stronger relationship between the \ac{DRR} and the learned representations compared to the \ac{T60}. This is notable because the \ac{T60} is generally considered to be a room-dependent parameter  in contrast to the \ac{DRR} that relates to the source-receiver distances \cite{Naylor2010b}.   This finding is in line with the observations in  \cite{Papayiannis2017},   which show that sub-band \ac{DRR} measurements are useful for the task of room classification when using the \ac{ACE} dataset. A relevant aspect of the \ac{ACE} dataset is the fact that the rooms are of the scale typically expected in residential and office buildings, which makes their \ac{T60} relatively short compared to larger rooms, such as the ones found in the \ac{QMUL} dataset (see Fig.~\ref{figure_rooms_t60_drr}). One interpretation of the above is that, for  spaces with similar characteristics to the ones in the \ac{ACE} dataset, the ratio between the energy of the strong early reflections  to the remaining energy can be more informative for room classification than  features that only capture the energy decay. Of course the learned representations are not bound to follow a strict definition of the \ac{DRR} or the \ac{T60}, which generally makes \acp{DNN} able to learn the most useful features for the task. This benefit sets \ac{DNN} classifiers apart from methods such as \cite{Moore2014}, where the representation of the input is fixed prior to training. Another important point to note here is that because of this benefit  the learned representations are to some degree ``tailored'' to the training dataset and using different datasets will lead to different representations. This is not a concern in settings where the training and testing rooms are fixed and known at the training time but might impact the usefulness of the learned representations with regards to transfer-learning to other tasks or to incremental-training settings, where rooms are incrementally added to the existing set.

The code implementation of the work discussed in this paper can be found at: \url{https://github.com/papayiannis/reverberation_learning_python}.

\ifCLASSOPTIONcaptionsoff
  \newpage
\fi

%\bibliographystyle{IEEEbibIn}
%\bibliography{sapstrings,papstrings,sapref}

\begin{thebibliography}{10}

\bibitem{Papayiannis2017}
C.~Papayiannis, C.~Evers, and P.~A. Naylor,
\newblock ``Discriminative feature domains for reverberant acoustic
  environments,''
\newblock in {\em Proc. {IEEE} Intl. Conf. on Acoustics, Speech and Signal
  Processing ({ICASSP})}, New Orleans, Louisiana, USA, Mar. 2017, pp. 756--760.

\bibitem{Moore2014}
A.~H. Moore, M.~Brookes, and P.~A. Naylor,
\newblock ``Room identification using roomprints,''
\newblock in {\em Proc. Audio Eng. Soc. ({AES}) Conf. on Audio Forensics}, June
  2014.

\bibitem{Peters2012}
N.~Peters, H.~Lei, and G.~Friedland,
\newblock ``Name that room: {{Room}} identification using acoustic features in
  a recording,''
\newblock in {\em Proceedings of the 20th {{ACM International Conference}} on
  {{Multimedia}}}, 2012, pp. 841--844.

\bibitem{Moore2018nsv}
A.~H. {Moore}, P.~A. {Naylor}, and M.~{Brookes},
\newblock ``Room identification using frequency dependence of spectral decay
  statistics,''
\newblock in {\em Proc. {IEEE} Intl. Conf. on Acoustics, Speech and Signal
  Processing ({ICASSP})}, 2018, pp. 6902--6906.

\bibitem{Eaton2015d}
J.~Eaton, N.~D. Gaubitch, A.~H. Moore, and P.~A. Naylor,
\newblock ``Proceeding of the {{ACE Challenge}},''
\newblock Proceedings, New Paltz, NY, USA, Oct. 2015.

\bibitem{PapayiannisThesis}
C.~Papayiannis,
\newblock {\em Models for learning reverberant environments},
\newblock Ph.D. thesis, Imperial College London, 2019.

\bibitem{Cox2001}
T.~J. Cox, F.~Li, and P.~Darlington,
\newblock ``Extracting room reverberation time from speech using artificial
  neural networks,''
\newblock {\em J. Audio Eng. Soc. ({AES})}, vol. 49, no. 4, pp. 219--230, 2001.

\bibitem{Xiong2019}
F.~{Xiong}, S.~{Goetze}, B.~{Kollmeier}, and B.~T. {Meyer},
\newblock ``Joint estimation of reverberation time and early-to-late
  reverberation ratio from single-channel speech signals,''
\newblock {\em IEEE/ACM Transactions on Audio, Speech, and Language
  Processing}, vol. 27, no. 2, pp. 255--267, 2019.

\bibitem{gamper2018blind}
H.~Gamper and I.~Tashev,
\newblock ``Blind reverberation time estimation using a convolutional neural
  network,''
\newblock in {\em Proc. International Workshop on Acoustic Signal Enhancement
  (IWAENC)}. September 2018, pp. 1--5, IEEE,
\newblock Nominated for best paper award.

\bibitem{Genovese2019}
A.~F. {Genovese}, H.~{Gamper}, V.~{Pulkki}, N.~{Raghuvanshi}, and I.~J.
  {Tashev},
\newblock ``Blind room volume estimation from single-channel noisy speech,''
\newblock in {\em Proc. {IEEE} Intl. Conf. on Acoustics, Speech and Signal
  Processing ({ICASSP})}, 2019, pp. 231--235.

\bibitem{Farina2000}
A.~Farina,
\newblock ``Simultaneous measurement of impulse response and distortion with a
  swept-sine technique,''
\newblock in {\em Proc. Audio Eng. Soc. ({AES}) Convention}, Feb. 2000, pp.
  1--23.

\bibitem{Sainath2017}
T.~Sainath, R.~J. Weiss, K.~Wilson, B.~Li, A.~Narayanan, E.~Variani,
  M.~Bacchiani, I.~Shafran, A.~Senior, K.~Chin, A.~Misra, and C.~Kim,
\newblock ``Multichannel {{Signal Processing}} with {{Deep Neural Networks}}
  for {{Automatic Speech Recognition}},''
\newblock {\em IEEE /ACM Transactions on Audio, Speech, and Language
  Processing}, vol. 25, pp. 965 -- 979, 2017.

\bibitem{Yan2019}
J.~Yan and Y.~Song,
\newblock ``Weakly labeled sound event detection with resdual {CRNN} using
  semi-supervised method,''
\newblock Tech. {R}ep., University of Science and Technology of China, National
  Engineering Laboratory for Speech and Language Information Processing, Hefei,
  China, June 2019.

\bibitem{Papayiannis2018}
C.~Papayiannis, J.~Amoh, V.~Rozgic, S.~Sundaram, and C.~Wang,
\newblock ``Detecting {{Media Sound Presence}} in {{Acoustic Scenes}},''
\newblock in {\em Proc. Conf. of Intl. Speech Commun. Assoc. ({INTERSPEECH})},
  Hyderabad, India, Sept. 2018, pp. 1363--1367.

\bibitem{Chan2019las}
W.~{Chan}, N.~{Jaitly}, Q.~{Le}, and O.~{Vinyals},
\newblock ``Listen, attend and spell: A neural network for large vocabulary
  conversational speech recognition,''
\newblock in {\em Proc. {IEEE} Intl. Conf. on Acoustics, Speech and Signal
  Processing ({ICASSP})}, 2016, pp. 4960--4964.

\bibitem{BahdanauCB14}
D.~Bahdanau, K.~Cho, and Y.~Bengio,
\newblock ``Neural machine translation by jointly learning to align and
  translate,''
\newblock in {\em Int. Conf. on Learning Representations {(ICLR)}}, 2015.

\bibitem{XuBKCCSZB15}
K.~Xu, J.~Ba, R.~Kiros, K.~Cho, A.~Courville, R.~Salakhudinov, R.~Zemel, and
  Y.~Bengio,
\newblock ``Show, attend and tell: Neural image caption generation with visual
  attention,''
\newblock in {\em Int. Conf. on Machine Learning}, 2015, pp. 2048--2057.

\bibitem{cnn_asr2014}
O.~{Abdel-Hamid}, A.~{Mohamed}, H.~{Jiang}, L.~{Deng}, G.~{Penn}, and D.~{Yu},
\newblock ``Convolutional neural networks for speech recognition,''
\newblock {\em IEEE/ACM Transactions on Audio, Speech, and Language
  Processing}, vol. 22, no. 10, pp. 1533--1545, 2014.

\bibitem{Simonyan2014}
K.~Simonyan and A.~Zisserman,
\newblock ``Very deep convolutional networks for large-scale image
  recognition,''
\newblock in {\em Int. Conf. on Learning Representations {(ICLR)}}, 2015.

\bibitem{Zhou1988}
Y.-T. Zhou, R.~Chellappa, A.~Vaid, and B.~K. Jenkins,
\newblock ``Image restoration using a neural network,''
\newblock {\em IEEE Transactions on Acoustics, Speech, and Signal Processing},
  vol. 36, no. 7, pp. 1141--1151, July 1988.

\bibitem{Hinton2012}
G.~E. Hinton, N.~Srivastava, A.~Krizhevsky, I.~Sutskever, and R.~Salakhutdinov,
\newblock ``Improving neural networks by preventing co-adaptation of feature
  detectors,''
\newblock {\em Computing Research Repository}, vol. abs/1207.0580, 2012.

\bibitem{Hahnloser2000}
R.~H.~R. Hahnloser, R.~Sarpeshkar, M.~A. Mahowald, R.~J. Douglas, and H.~S.
  Seung,
\newblock ``Digital selection and analogue amplification coexist in a
  cortex-inspired silicon circuit,''
\newblock {\em Nature}, vol. 405, pp. 947, June 2000.

\bibitem{Sainath2015}
T.~N. Sainath, O.~Vinyals, A.~Senior, and H.~Sak,
\newblock ``Convolutional, {{Long Short}}-{{Term Memory}}, fully connected
  {{Deep Neural Networks}},''
\newblock in {\em Proc. {IEEE} Intl. Conf. on Acoustics, Speech and Signal
  Processing ({ICASSP})}, Brisbane, Australia, Apr. 2015, pp. 4580--4584.

\bibitem{Garofolo1993}
J.~S. Garofolo, L.~F. Lamel, W.~M. Fisher, J.~G. Fiscus, D.~S. Pallett, N.~L.
  Dahlgren, and V.~Zue,
\newblock ``{{TIMIT}} acoustic-phonetic continuous speech corpus,''
\newblock Corpus, Linguistic Data Consortium ({LDC}), Philadelphia, 1993.

\bibitem{Eaton2015a}
J.~Eaton, N.~D. Gaubitch, A.~H. Moore, and P.~A. Naylor,
\newblock ``The {{ACE Challenge}} - corpus description and performance
  evaluation,''
\newblock in {\em Proc. {IEEE} Workshop on Applications of Signal Processing to
  Audio and Acoustics ({WASPAA})}, New Paltz, NY, USA, 2015.

\bibitem{Stewart2010}
R.~Stewart and M.~Sandler,
\newblock ``Database of omnidirectional and {{B}}-format room impulse
  responses,''
\newblock in {\em Proc. {IEEE} Intl. Conf. on Acoustics, Speech and Signal
  Processing ({ICASSP})}, Dallas, Texas, USA, Mar. 2010, pp. 165--168.

\bibitem{Kingma2014}
D.~P. Kingma and J.~Ba,
\newblock ``Adam: {A} method for stochastic optimization,''
\newblock in {\em Int. Conf. on Learning Representations {(ICLR)}}, 2015.

\bibitem{lin_ccc}
L.~I.-K. Lin,
\newblock ``A concordance correlation coefficient to evaluate
  reproducibility,''
\newblock {\em Biometrics}, vol. 45, no. 1, pp. 255--268, 1989.

\bibitem{spectral_bandwidth}
P.~Herrera-Boyer, A.~Klapuri, and M.~Davy,
\newblock {\em Automatic Classification of Pitched Musical Instrument Sounds},
  pp. 163--200,
\newblock Springer US, Boston, MA, 2006.

\bibitem{spectral_rolloff}
G.~Peeters,
\newblock ``A large set of audio features for sound description (similarity and
  classification) in the cuidado project,''
\newblock {\em CUIDADO IST Project Report}, vol. 54, no. 0, pp. 1--25, 2004.

\bibitem{Naylor2010b}
P.~A. Naylor and N.~D. Gaubitch, Eds.,
\newblock {\em Speech Dereverberation},
\newblock Springer, 2010.

\end{thebibliography}

\end{document}